\documentclass[10pt,aps,prd,nofootinbib,superscriptaddress,twocolumn,preprintnumbers,balancelastpage]{revtex4}

\usepackage{graphicx,epsfig,psfrag,amssymb,hyperref}
\usepackage{multirow}
\usepackage{color,graphicx,epsfig,psfrag,amsmath,empheq}
\usepackage{bm}
\usepackage{mathrsfs,amsfonts,color}
\usepackage{slashed}
\usepackage{hepunits}
\usepackage{color}
\usepackage{enumitem}

\newcommand{\cL}{\mathcal{L}}

\newcommand{\cM}{\mathcal{M}}
\newcommand{\cR}{\mathcal{R}}

\newcommand{\cP}{\mathcal{P}}

\newcommand{\cH}{\mathcal{H}}

\newcommand{\abs}[1]{\left\lvert#1\right\rvert}

\definecolor{greenp1}{rgb}{0, 0.8, 0}
\definecolor{danielColor}{rgb}{0.9, 0.2, 0.9}

\newcommand{\primex}{{\sc PrimEx}\xspace}
\newcommand{\gluex}{{\sc GlueX}\xspace}

\begin{document}

\title{Photoproduction of axion-like particles}

\author{Daniel Aloni}
\email{daniel.aloni@weizmann.ac.il}
\affiliation{Department of Particle Physics and Astrophysics, Weizmann Institute of Science, Rehovot, Israel 7610001}

\author{Cristiano Fanelli}
\email{cfanelli@mit.edu}
\affiliation{Laboratory for Nuclear Science, Massachusetts Institute of Technology, Cambridge, MA 02139, U.S.A.}

\author{Yotam Soreq}
\email{yotam.soreq@cern.ch}
\affiliation{Theoretical Physics Department, CERN, CH-1211 Geneva 23, Switzerland}
\affiliation{Department of Physics, Technion, Haifa 32000, Israel}

\author{Mike Williams}
\email{mwill@mit.edu}
\affiliation{Laboratory for Nuclear Science, Massachusetts Institute of Technology, Cambridge, MA 02139, U.S.A.}

\begin{abstract}
We explore the sensitivity of photon-beam experiments to axion-like particles~(ALPs) with QCD-scale masses whose dominant coupling to the Standard Model is either to photons or gluons.
We introduce a novel data-driven method that eliminates the need for knowledge of nuclear form factors or the photon-beam flux when considering coherent Primakoff production off a nuclear target, and show that data collected by the \primex experiment could substantially improve the sensitivity to ALPs with $0.03 \lesssim m_a \lesssim 0.3$\,GeV.
Furthermore, we explore the potential sensitivity of running the \gluex experiment with a nuclear target and its planned \primex-like calorimeter.
For the case where the dominant coupling is to gluons, we study photoproduction for the first time, and predict the future sensitivity of the \gluex experiment using its nominal proton target.
Finally, we set world-leading limits for both the ALP-gluon coupling and the ALP-photon coupling based on public mass plots.
\end{abstract}

\preprint{CERN-TH-2019-023}

\maketitle

Axion-like particles~(ALPs) are hypothetical pseudoscalars found in many proposed extensions to the Standard Model~(SM), since they naturally address the Strong $CP$\,\cite{Peccei:1977hh,Peccei:1977ur,Weinberg:1977ma,Wilczek:1977pj} and Hierarchy problems~\cite{Graham:2015cka}.
Furthermore, ALPs may explain the muon magnetic moment anomaly~\cite{Chang:2000ii,Marciano:2016yhf},
and could connect SM particles to dark matter by providing a {\em portal}~\cite{Nomura:2008ru,Freytsis:2010ne,Dolan:2014ska,Hochberg:2018rjs}.
The couplings of ALPs to the SM are highly suppressed at low energies by a large cut-off scale $\Lambda$; however, since ALPs, $a$, are pseudo-Nambu-Goldstone bosons, their mass~$(m_a)$ can be much smaller than the scale that controls their dynamics, {\em i.e.}\ $m_a\ll \Lambda\,$.
Recently, ALPs with MeV-to-GeV scale masses, henceforth QCD scale, have received considerable interest~\cite{Marciano:2016yhf,Jaeckel:2015jla,Dobrich:2015jyk,Izaguirre:2016dfi,Knapen:2016moh,Mariotti:2017vtv,Bauer:2017ris,CidVidal:2018blh,Bauer:2018uxu,Harland-Lang:2019zur,Ebadi:2019gij,Mimasu:2014nea,Brivio:2017ije}
(see, in addition, Refs.\cite{Essig:2013lka,Marsh:2015xka,Graham:2015ouw,Irastorza:2018dyq,Beacham:2019nyx} for recent ALP reviews).

\begin{figure}[t]
\includegraphics[width=0.27\textwidth]{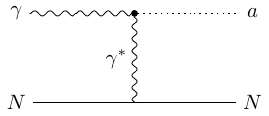}
\includegraphics[width=0.27\textwidth]{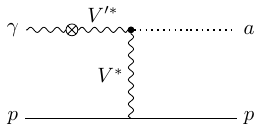}
\caption{
(top) Primakoff production via $t$-channel photon exchange, and
(bottom) photoproduction via photon--vector-meson mixing and $t$-channel vector-meson exchange.
}
\label{fig:feyn}
\end{figure}

In this Letter, we explore the discovery potential of photon-beam experiments for ALPs with QCD-scale masses.
Specifically, we consider two cases: ALPs whose dominant coupling to SM particles is to photons or to gluons.
For the former, the best sensitivity involves coherent Primakoff production off a nuclear target (see Fig.~\ref{fig:feyn}~top).
While ALP production using the Primakoff process has been studied before~\cite{GasparianTalk,Marciano:2016yhf}, our work is novel in three aspects:
(i)~we introduce a fully data-driven ALP normalization method, which eliminates the need for knowledge of nuclear form factors or the photon-beam flux;
(ii)~we show that data collected by the \primex experiment at Jefferson Lab could substantially improve the sensitivity to ALPs with $0.03 \lesssim m_a \lesssim 0.3$\,GeV, in fact, we are able to set world-leading limits from a diphoton mass plot published in Ref.~\cite{Larin:2010kq} from a single angular bin;
and (iii)~we explore for the first time the potential sensitivity of running the \gluex experiment at Jefferson Lab with a nuclear target and its planned \primex-like calorimeter.
For the case where the dominant SM coupling of ALPs is to gluons, we extend our work in Ref.~\cite{Aloni:2018vki} and study photoproduction for the first time.
The dominant photoproduction mechanism is photon--vector-meson mixing and $t$-channel vector-meson exchange (see Fig.~\ref{fig:feyn}~bottom).
We obtain the future sensitivity of the \gluex experiment using its nominal proton target, and set world-leading limits based on a public mass plot.

The effective Lagrangian describing the interactions of ALPs with photons and gluons is
\begin{align}
	\label{eq:Leff}
	\cL_{\rm eff}
	\supset
	 \frac{c_\gamma}{4\Lambda} a F^{\mu\nu}\tilde{F}_{\mu\nu}
  	- \frac{4\pi \alpha_s c_g }{\Lambda} a G^{\mu\nu}\tilde{G}_{\mu\nu} \, ,
\end{align}
where $F_{\mu\nu}\,(G_{\mu\nu})$ is the photon\,(gluon) field strength tensor with $\tilde{F}_{\mu\nu} =\frac{1}{2}\epsilon_{\mu\nu\alpha\beta}F^{\alpha\beta}$ ($\tilde{G}_{\mu\nu}$ satisfies a similar expression).
Our approach to studying ALP-hadron interactions follows Refs.~\cite{Georgi:1986df,Bardeen:1986yb,Krauss:1986bq}, and we take the ALP-pseudoscalar mixing, along with the ALP lifetime and branching fractions, directly from Ref.~\cite{Aloni:2018vki}.
The two scenarios considered in this Letter correspond to
$c_g\!=\!0$, $c_\gamma\!=\!1$ and $c_g\!=\!1$, $c_\gamma\!=\!0$; however, we stress that our results can be generalized to any other set of ALP couplings to the SM particles (see Ref.~\cite{Aloni:2018vki}).

First, we consider the case where the dominant ALP-SM coupling is to photons.
When a photon beam is incident on a nuclear target, the production of pseudoscalars---either the mesons $P = \pi^0, \eta$ or ALPs---at forward angles is dominantly via the coherent Primakoff process for $m_{a,P} \lesssim 1$\,GeV.
The differential cross section for elastic coherent Primakoff production from a nucleus, $N$, is given by
\begin{align}
	\label{eq:dsigdt}
	\frac{ d \sigma_{\gamma N \to a N}^{\rm elastic} }{ d t }
\!=\!	\alpha Z^2_N F^2_{N}(t)
	\Gamma_{a\to\gamma\gamma}
	 \cH(m_N, m_a, \,s, \, t)
	 \, ,
\end{align}
where $t$ and $s$ are the Mandelstam variables,
$F_N$ is the nuclear form factor (see the Supplemental Material~\cite{SuppMat} to this Letter and Refs.~\cite{RevModPhys.29.144,PhysRevLett.8.110,PDG}),
$\Gamma_{a\to\gamma\gamma}=c^2_\gamma m^3_a/(64\pi\Lambda^2)$ is the partial decay width of the decay $a \to \gamma\gamma$,
and
\begin{align}
&	\cH(m_N, m_a, \,s, \, t)
	\equiv 128\pi\frac{m^4_N}{m^3_a}   \nonumber\\
&
	\times \frac{m_a^2 t(m_N^2+s) - m_a^4 m_N^2 - t( (s-m^2_N)^2 + st ) }{t^2(s-m^2_N)^2(t-4m^2_N)^2} \, .
\end{align}
For pseudoscalar mesons, the corresponding differential cross section is obtained by the replacement  $a \to P$.

For small values of $t$ (forward angles), where elastic coherent Primakoff production is dominant, the nuclear form factor dependence can be canceled by forming the ratio of the ALP and $P$ differential cross sections as follows:
\begin{align}
	\label{eq:dsigdtDataDriven}
	\frac{ d \sigma_{\! \gamma N \to a N}^{\rm elastic} }{ d t }
=	\frac{\Gamma_{a\to\gamma\gamma}}{\Gamma_{P\to\gamma\gamma}}
	\frac{\cH(m_N, m_a, \, s, \, t)}{\cH(m_N, m_P, \, s, \, t)}
	\frac{ d \sigma_{\!\gamma N \to P N}^{ \rm elastic} }{ d t }\, .
\end{align}
Therefore, the ALP yield---up to a factor of the model parameters $(c_{\gamma}/\Lambda)^2$---can be determined from the observed $\pi^0$ and/or $\eta$ Primakoff yields,
making this a completely data-driven search.
The nuclear form factor does not need to be known, and the photon flux also cancels using our approach.
A correction must be applied to account for any mass dependence in the detector efficiency at fixed $t$ and $s$, though this should be easy to obtain from simulation given that the $a\to\gamma\gamma$ decay distribution is known (it must be uniform in the $a$ rest frame).
Finally, we note that quasi-elastic ALP Primakoff production can be estimated using a similar approach; however, this production mechanism is negligible in the $m_a$ range considered here (see the Supplemental Material~\cite{SuppMat}).

The first run of the \primex experiment was in Hall~B at Jefferson Lab in 2004~\cite{Larin:2010kq}.
Data were collected on both C and Pb targets using a 4.9--5.5\,GeV photon beam and a high-resolution multichannel calorimeter, which allowed \primex to make the most precise measurement to date of the $\pi^0 \to \gamma\gamma$ decay width.
The integrated luminosities were 0.3/pb for C and 0.002/pb for Pb.
A follow-up run of \primex was performed in 2010, which collected 0.4/pb on C and 0.3/pb on Si, though only preliminary results have been produced thus far from this data set.

\begin{figure}[!t]
\includegraphics[width=0.49\textwidth]{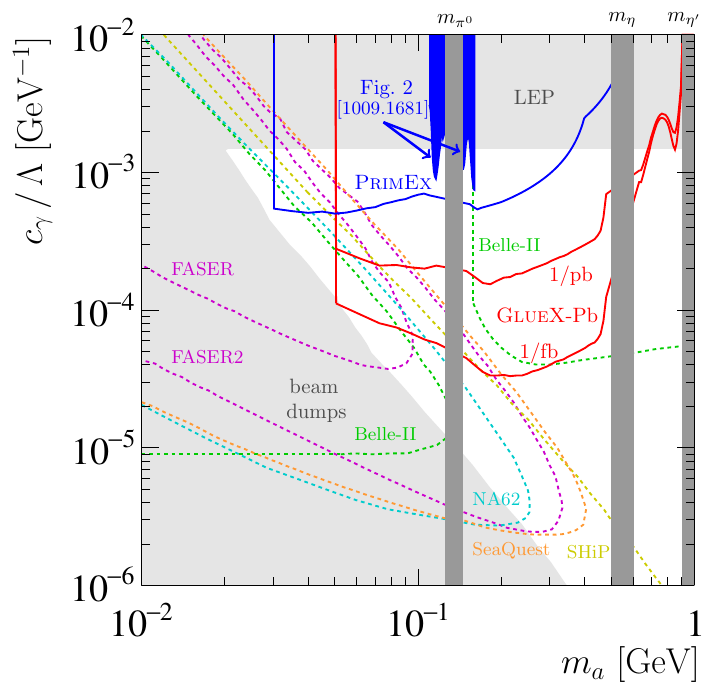}
\caption{
The \primex\,(blue) and \gluex\,(red) projections for the ALP-photon coupling ($c_\gamma=1$, $c_g=0$) compared to the current bounds~\cite{Bjorken:1988as,Blumlein:1990ay,Abbiendi:2002je,Knapen:2016moh} and projections of NA62, SeaQuest, Belle~2, SHiP and FASER~\cite{Feng:2018noy,Berlin:2018pwi,Dobrich:2015jyk,Dolan:2017osp}.
In addition, a new limit is set (dark blue shaded regions) using the published $m_{\gamma\gamma}$ spectrum from one angular bin of carbon-target \primex data from Fig.~2 of Ref.~\cite{Larin:2010kq}.
 }
\label{fig:prim_lims}
\end{figure}

Reference~\cite{Larin:2010kq} published the diphoton mass spectrum near the $\pi^0$ peak for one forward angular bin from the C data obtained in the first \primex run (see Fig.~2 of Ref.~\cite{Larin:2010kq}).
The diphoton efficiency is roughly constant within such a small angular and mass window; therefore, using the observed $\pi^0$ yield in the published peak (${\approx 5100}$) and the background yield at each $m_{\gamma\gamma}$, we can use Eq.~\eqref{eq:dsigdtDataDriven} to place constraints on $\Lambda$ for $c_{\gamma}=1$ and $c_g = 0$.
For example, at $m_a = 0.11$\,GeV the background in a $\pm2\sigma$ window is $\approx 300$ giving a rough estimate of the sensitivity to the ALP yield of $\approx 2\sqrt{300}$.
Using Eq.~\eqref{eq:dsigdtDataDriven} we estimate this corresponds to $\Lambda \approx 0.6$\,TeV, which is comparable to the world-leading constraint from LEP at this mass~\cite{Abbiendi:2002je,Knapen:2016moh}.
In the Supplemental Material~\cite{SuppMat}, we perform a more rigorous study of this spectrum, the results of which are shown in Fig.~\ref{fig:prim_lims} and confirm that this small fraction of the \primex data sample provides competitive sensitivity to LEP---and even gives world-leading constraints at a few masses.

To estimate the sensitivity of each \primex data sample, {\em i.e.}\ not just the one bin shown in Fig.~2 of Ref.~\cite{Larin:2010kq}, we need to determine the mass dependence of the efficiency and to estimate the background versus $m_{\gamma\gamma}$ in each sample.
A detailed description of this part of the analysis is provided in the Supplemental Material~\cite{SuppMat}, and briefly summarized here.
We assume that the same selection criteria applied in Ref.~\cite{Larin:2010kq} are used for the ALP search and take the \primex calorimeter acceptance and resolution from Refs.~\cite{kubantsev2006performance,larin2011new}.
Furthermore, we assume that the ALP bump hunt will only use candidates with $\theta_{\gamma\gamma} < 0.5^{\circ}$, where $\pi^0$ production is dominated by the Primakoff process for all targets.

Using the known nuclear form factors and Primakoff differential cross section~\cite{Donnelly:2017aaa,Alberico:1988bv,DeJager:1974liz}, we generate Primakoff $\pi^0$ Monte Carlo events for the \primex photon-beam energy.
We require that both photons from the $\pi^0\to\gamma\gamma$ decay are in the \primex calorimeter fiducial acceptance region~\cite{Larin:2010kq} and apply the required smearing to account for resolution.
The width of our Monte Carlo $\pi^0$ peak is consistent with the data in Ref.~\cite{Larin:2010kq}.
We then apply the full selection of Ref.~\cite{Larin:2010kq} and find that our predicted $\pi^0$ Primakoff yields are consistent with those observed in Refs.~\cite{PrimExPAC33,YangThesis}.
We assume that the reconstruction efficiency is independent of the ALP mass in the search region (excluding geometrical acceptance effects).
In addition, we discard the low-$m_a$ region where the photon clusters begin to overlap and the acceptance has strong mass dependence.
Lower masses can likely be explored in an analysis of the actual \primex data with access to a full detector simulation.

As in any bump hunt, obtaining a data-driven background estimate is straightforward using the $m_{\gamma\gamma}$ sidebands at each $m_a$ (see, {\em e.g.}, Refs.~\cite{Williams:2015xfa,Williams:2017gwf}).
However, estimating the background without the data is considerably more difficult, so we adopt a conservative approach.
We considered many possible backgrounds, {\em e.g.}\ $\gamma N \to N \omega(\pi^0[\gamma\gamma]\gamma)$ where one photon is not reconstructed or the $\pi^0$ photons are merged into a single cluster, though we found that no hadronic reactions are capable of contributing background at a rate comparable to that observed in Fig.~2 of Ref.~\cite{Larin:2010kq} in the mass range probed by \primex.
Thus, we conclude that the \primex background is dominantly due to electromagnetic interactions of the photon beam with the target that produce either additional photons or $e^+e^-$ pairs.
Figure~2 of Ref.~\cite{Larin:2010kq} shows the forward-most angular region.
Given that the beam backgrounds should decrease moving away from the beam line, using this angular bin---and assuming a uniformly distributed background---provides a conservative background estimate.
We also conservatively assume that the background density above (below) the $m_{\gamma\gamma}$ region shown in Fig.~2 of Ref.~\cite{Larin:2010kq} takes on the value at the upper-most (lower-most) bin of the published $m_{\gamma\gamma}$  spectrum.
Finally, we scale the beam-induced background, which is shown for the first C run, by the product of the target radiation length and the number of photons on target for other \primex runs.

Our projected sensitivity for the entire \primex data sample is shown in Fig.~\ref{fig:prim_lims}, {\em i.e.}\ our estimate combines all of the \primex runs.
The \primex sensitivity would be substantially better than LEP for $0.03 \lesssim m_a \lesssim 0.3$\,GeV and provide world-leading sensitivity up to about 0.4\,GeV.
We stress again that the \primex data are already on tape, and are well calibrated and understood.
All that is needed is to perform a bump hunt on the forward-angle data in the region dominated by Primakoff production.
Following the approach we proposed above, the normalization can be done in a purely data-driven way using the observed $\pi^0$ Primakoff yield differentially versus $t$.

An updated version of the \primex experiment recently ran in Hall~D at Jefferson Lab using the \gluex detector with an additional small-angle calorimeter~\cite{Primakoff2010}.
This new experiment has the potential to explore higher masses than \primex due to the higher photon-beam energy of 11\,GeV and the larger acceptance of the \gluex forward calorimeter; however, the use of a helium target in this new run makes it less sensitive than \primex for ALPs.
There are several proposals for future \gluex running with heavy nuclear targets\,\cite{PAC}, so it is interesting to explore the potential sensitivity to ALPs of such runs.
Specifically, we consider a Pb target here.
We take the \gluex acceptance, efficiency, and resolution from Refs.~\cite{beattie2018construction,hardin2018upgrading}, and the corresponding values for the small-angle calorimeter from Ref.~\cite{Primakoff2010}.
For $m_a < m_{\eta}$, we rescale the expected beam background from the \primex Pb run.
There are three additional backgrounds that contribute to the \gluex run at higher masses:
Primakoff production of $\eta$ and $\eta'$ mesons, and
coherent nuclear production of $\gamma N \to N \omega(\pi^0[\gamma\gamma]\gamma)$ (as described above).
The cross sections for these processes are well known, making it straightforward to estimate their yields using Monte Carlo.

An additional complication arises when projecting the sensitivity of \gluex.
The \gluex experiment could explore regions of ALP parameter space where the ALP flight distance becomes nonnegligible.
Using Monte Carlo, we estimate that the impact on the ALP mass resolution and acceptance is small provided that its lab-frame flight distance is $\lesssim 30$\,cm (the length of the nominal liquid hydrogen target cell).
For simplicity, we apply a fiducial cut on the flight distance at 30\,cm, which is conservative since ALPs that decay after this distance could still be detected and a detailed study could determine the appropriate signal shape for each value of $\Lambda$.
Our estimate of the projected reach for \gluex including the \primex-like calorimeter is shown in Fig.~\ref{fig:prim_lims}.
The larger data set assumes that as much data is collected as is expected in the full \gluex proton-target run.
The smaller data set corresponds to collecting 1/pb of Pb-target data.

Figure~\ref{fig:prim_lims} shows that Primakoff production using photon beams can provide unique sensitivity to ALPs.
The data from \primex, which has been on tape for a decade, could provide substantially better sensitivity than LEP for $0.03 \lesssim m_a \lesssim 0.3$\,GeV.
Running the \gluex experiment for a few years with a Pb target and its \primex-like small-angle calorimeter could explore the remaining parameter space down to where future beam-dump experiments will have sensitivity.
Much of this parameter space is not accessible at any other current or proposed future experiment.

We now move on to considering the case where the dominant ALP-SM coupling is to gluons.
In this scenario, nuclear targets do not provide a large advantage since both the signal and background scale similarly with the number of nucleons, so we consider the nominal liquid-hydrogen target and default experimental \gluex setup, which has been running for the past few years.
When a photon beam is incident on a proton target, exclusive pseudoscalar production is dominantly via photon--vector-meson mixing as shown in Fig.~\ref{fig:feyn}~bottom.
In the Supplemental Material~\cite{SuppMat}, we show that---once both $\pi^0$ and $\eta$ photoproduction are well understood---it is possible to derive a fully data-driven normalization strategy similar to the one we proposed above for Primakoff production.
As discussed in Ref.~\cite{AlGhoul:2017nbp}, $\eta$ production at \gluex energies, while clearly dominantly $t$-channel, is not yet fully understood.
Therefore, we will adopt a simplified approach below, though we do provide a complete description of how to implement the fully data-driven strategy for future searches in the Supplemental Material~\cite{SuppMat}.

In principle, ALP searches at \gluex could look for hadronic final states like $a \to 3\pi$ and $a\to \eta\pi\pi$; however, we studied these and found that the mass resolution at \gluex makes ALP peaks comparable in width to $\omega, \eta', \phi \to 3\pi$ and $\eta', f_2 \to \eta\pi\pi$ making it likely that large mass regions need to be vetoed in such searches.
Furthermore, the sensitivity at higher masses would not be competitive with $b$\,-hadron decays~\cite{Aloni:2018vki}.
Therefore, we choose to focus on the $a\to\gamma\gamma$ decay in the region between the $\pi^0$ and $\eta$ mesons, where its branching fraction is close to unity and diphoton backgrounds are small.
Since $a \to \pi\pi$ and $a \to \pi^0\gamma$ are forbidden by $CP$ and $C$, respectively, the $a\to\gamma\gamma$ decay is dominant in all ALP models in most of this mass region.

For $m_{\pi^0} < m_a < m_{\eta}$, the ALP-gluon coupling can be replaced by ALP--pseudoscalar-meson mixing by performing a chiral transformation of the light-quark fields\,\cite{Georgi:1986df,Bardeen:1986yb,Krauss:1986bq} .
Following Ref.~\cite{Aloni:2018vki}, we denote the mixing of the ALP with the $\pi^0$ and $\eta$ as
$\langle \boldsymbol{a \pi^0} \rangle$ and $\langle \boldsymbol{a \eta} \rangle$, respectively, and we take these $m_a$-dependent mixings directly from Ref.~\cite{Aloni:2018vki}.
For $|t| \lesssim 1$\,GeV$^2$ in this $m_a$ region, at fixed $s$ and $t$ the following approximation is valid to $\mathcal{O}(1)$, which is roughly the same fidelity with which the ALP-pseudoscalar mixing terms are known:
\begin{align}
	\label{eq:strongnorm}
	\frac{ d \sigma_{\! \gamma p \to a p}}{ d t }
	\approx
& 	\left( \frac{f_{\pi}}{f_a} \right)^2  \\
& 	\times \left[|\langle \boldsymbol{a \pi^0} \rangle|^2 	\frac{ d \sigma_{\! \gamma p \to \pi^0 p} }{ d t }
	+ |\langle \boldsymbol{a \eta} \rangle|^2 	\frac{ d \sigma_{\! \gamma p \to \eta p} }{ d t }   \right], \nonumber
\end{align}
where $f_{\pi}$ and $f_a = \Lambda / (32 \pi^2 c_g)$ are the pion and ALP decay constants.
This approximation works well in this mass range due to the dominance of the contributions from $\pi^0$ or $\eta$ mixing to the ALP $U(3)$ representation.
We adopt the relevant numerical values from Refs.~\cite{Fujiwara:1984mp,Machleidt:1987hj,Mathieu:2015eia}, see Supplemental Material for details.

\begin{figure}[t]
\includegraphics[width=0.49\textwidth]{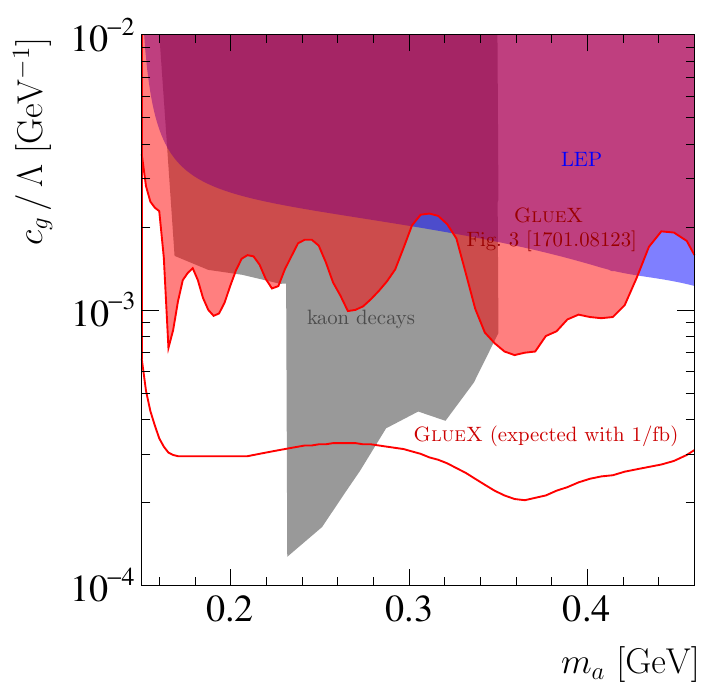}
\caption{
The \gluex projection for the ALP-gluon coupling ($c_\gamma=0$, $c_g=1$) compared to the current bounds~\cite{Aloni:2018vki} from LEP~\cite{Abbiendi:2002je,Knapen:2016moh} and kaon decays~\cite{Izaguirre:2016dfi,Ceccucci:2014oza,Abouzaid:2008xm,TobiKpLim}.
In addition, a new limit is set using the published $m_{\gamma\gamma}$ spectrum from $\approx 1$/pb of \gluex data from Fig.~3 of Ref.~\cite{AlGhoul:2017nbp}.
 }
 \begin{flushleft}
\end{flushleft}
\label{fig:strong_lims}
\end{figure}

Ref.~\cite{AlGhoul:2017nbp} published the $m_{\gamma\gamma}$ spectrum, along with the yields and efficiencies versus $t$ of both the $\pi^0$ and $\eta$ mesons.
In the Supplemental Material, we perform a bump hunt of the $m_{\gamma\gamma}$ spectrum to obtain upper limits on the ALP yield at each $m_a$.
The expected ALP yield in a small bin of $[s,t]$ is related to $\Lambda$ (or $f_a$) using Eq.~\eqref{eq:strongnorm} according to
\begin{align}
	\label{eq:strongyield}
	n_a(s,t)
	\approx
& 	\left( \frac{f_{\pi}}{f_a} \right)^2
	 \Bigg[ |\langle \boldsymbol{a \pi^0} \rangle|^2 \frac{n_{\pi^0}(s,t)\epsilon(m_a,s,t)}{\mathcal{B}(\pi^0\to\gamma\gamma)\epsilon(m_{\pi},s,t)}  \\
& 	+  |\langle \boldsymbol{a \eta} \rangle|^2 \frac{ n_{\eta}(s,t) \epsilon(m_a,s,t)}{\mathcal{B}(\eta \to \gamma\gamma) \epsilon(m_{\eta},s,t)}  \Bigg] \mathcal{B}(a \to \gamma\gamma) \,, \nonumber
\end{align}
where $\epsilon$ denotes the product of the detector acceptance and efficiency.
We linearly interpolate the efficiencies given in Ref.~\cite{AlGhoul:2017nbp} at $m_{\pi^0}$ and $m_{\eta}$ for $m_a$, and confirm this approach is valid to $\mathcal{O}(10\%)$ using toy Monte Carlo as described in the Supplemental Material~\cite{SuppMat} (additionally, the same ALP lifetime correction is applied here, though this is a small correction).
The total expected ALP yield is simply the sum of $n_a(s,t)$ over all bins.
By comparing the expected ALP yields to the upper limits obtained from the bump hunt of the $m_{\gamma\gamma}$ spectrum, we place constraints on $c_g/\Lambda$
(see Fig.~\ref{fig:strong_lims}).
These limits are the best over some of the $0.15 < m_a < 0.46$\,GeV region.
Finally, we also provide the expected sensitivity from a 1/fb \gluex data set, which is substantially better than any existing limits over most of this mass region.

In summary,
we explored the sensitivity of photon-beam experiments to ALPs with QCD-scale masses whose dominant coupling is to either photons or gluons.
For the photon-dominant coupling scenario,
we introduced a novel data-driven method that eliminates the need for knowledge of nuclear form factors or the photon-beam flux when considering coherent Primakoff production off of a nuclear target,
and showed that data collected by \primex could substantially improve the sensitivity to ALPs with $0.03 \lesssim m_a \lesssim 0.3$\,GeV.
We also explored the potential sensitivity of running the \gluex experiment with a nuclear target.
For the case where the dominant coupling is to gluons, we studied photoproduction for the first time,
and predicted the future sensitivity of the \gluex experiment using its nominal proton target.
For both scenarios, we set world-leading limits based on public mass plots.

\begin{acknowledgments}
We thank Bill Donnelly, Liping Gan, Ashot Gasparian, Or Hen, Ilya Larin, and Michael Spannowsky for useful discussions,
as well as Iftah Galon and Sebastian Trojanowski for providing us the details used in Fig.~\ref{fig:prim_lims} regarding other experiments,
and Gilad Perez and Kohsaku Tobioka for pointing out Ref.~\cite{TobiKpLim}.
YS and MW performed part of this work at the Aspen Center for Physics, which is supported by U.S.\ National Science Foundation grant PHY-1607611.
This work was supported by:
CS and MW were supported by the Office of Nuclear Physics of the U.S.\ Department of Energy under grant contract number DE-FG02-94ER40818;
and MW was also supported by the U.S.\ National Science Foundation under contract number PHY-1607225.
\end{acknowledgments}


\twocolumngrid
\vspace{-8pt}
\section*{References}
\vspace{-10pt}
\def\bibsection{}
\bibliographystyle{utphys}
\bibliography{a-gamma-gamma_bib}

\clearpage
\newpage
\maketitle
\onecolumngrid

\begin{center}
\textbf{\large Photoproduction of axion-like particles} \\
\vspace{0.05in}
{ \it \large Supplemental Material}\\
\vspace{0.05in}
{Daniel Aloni, Cristiano Fanelli, Yotam Soreq, and Mike Williams}
\end{center}

\onecolumngrid
\setcounter{equation}{0}
\setcounter{figure}{0}
\setcounter{table}{0}
\setcounter{section}{0}
\setcounter{page}{1}
\makeatletter
\renewcommand{\theequation}{S\arabic{equation}}
\renewcommand{\thefigure}{S\arabic{figure}}
\renewcommand{\thetable}{S\arabic{table}}
\newcommand\ptwiddle[1]{\mathord{\mathop{#1}\limits^{\scriptscriptstyle(\sim)}}}

\section{ALP Primakoff production}

This section considers the ALP Primakoff production process shown in Fig.~\ref{fig:feyn} (top)
\begin{align}
	\gamma(k_\gamma) + N (k_N) \to a(p_a) + X (p_X) \, ,
\end{align}
where $N$ is a nucleus at rest and $X$ is a generic final state with invariant mass $m_X$. We begin by defining the kinematical variables, and then explore the quasi-elastic~(QE) production on a heavy nuclear target, and the limiting cases of elastic production on both a heavy nuclear target and on a proton target.

\subsection{Kinematics }

The momenta in the lab frame are
\begin{align}
	\label{eq:k}
	& 	k_\gamma = (k, 0, 0, k) \, , \\
	&	k_N = (m_N, 0, 0, 0) \, , \\
	& 	p_a = \left( p, \sin\theta \sqrt{ p^2- m^2_a }, 0, \cos\theta \sqrt{ p^2- m^2_a }\right) \, , \\
	&	p_X = k_N + k_\gamma - p_a \, ,
\end{align}
where $\theta$ is the scattering angle.
The transferred momentum is
\begin{align}
	& 	q  =k_\gamma - p_a = (\omega, \ \vec{q}) \, .
\end{align}
We use the standard Mandelstam variables
\begin{align}
	\label{eq:s}
	&	s
	= 	(k_\gamma + k_N)^2
	= 	(p_a + p_X)^2 = m_N^2 + 2 m_N k    \, , \\
	\label{eq:t}
	&	t
	= 	(k_\gamma - p_a)^2
	= 	(k_N - p_X ) ^2 \, , \\
	\label{eq:u}
	&	u
	=	(k_\gamma - p_X)^2
	= 	(p_a - k_N ) ^2 \, ,
\end{align}
with $u+t+s = m_a^2 + m_N^2 + m_X^2\,$.
By using the above definitions we get that
\begin{align}
	m_X^2 &=  (k_N + q)^2 = m^2_N + 2m_N \omega + t  \ge m_N^2\, .
\end{align}
The elastic limit is $m_X^2=m_N^2$.

\subsection{Quasi-elastic double differential cross section}
We start from the most general case of Primakoff quasi-elastic production on a heavy nucleus.
The generic structure of the amplitude can be written as
\begin{align}
	i\cM
	= 	e \frac{ c_{\gamma}}{\Lambda } \frac{1}{t} J_{\rm had}^\mu J_{\gamma-a, \mu} \, ,
\end{align}
where $J_{\rm had}^\mu \,(J_{\gamma-a, \mu}) $ is the hadronic ($\gamma$--ALP) current.
The spin averaged squared amplitude is given by
\begin{align}
	\abs{\overline{\cM}}^2
	=	e^2 \frac{ c^2_{\gamma}}{\Lambda^2 } \frac{1}{t^2} L_{\mu\nu} W^{\mu\nu} \, .
\end{align}
Above, we adopt the standard parameterization of the hadronic tensor (see, for example, Ref.~\cite{Donnelly:2017aaa}),
\begin{align}
	\label{eq:Hadronic tensor}
	W^{\mu\nu}
	= 	\sum_{\rm had,avg} J^{\mu*}_{\rm had} J^\nu_{\rm had}
	=	-\widetilde{W}_1\left(g^{\mu\nu} - \frac{q^\mu q^\nu}{t} \right) + \widetilde{W}_2 V_i^\mu V_i^\nu \, ,
\end{align}
where the response functions $\widetilde{W}_1\,(\widetilde{W}_2)$ are only functions of $t$ and $x=-\frac{t}{2q \cdot k_N}$, the Bjorken parameter (or, alternatively, $m_X^2$), and
\begin{align}
	V_i^\mu
	=	\frac{1}{m_N}\left[ k_N^\mu - \frac{q \cdot k_N}{t} q^\mu  \right]
	=	\frac{1}{m_N}\left[ k_N^\mu + \frac{q^\mu}{2x} \right] \, .
\end{align}
The ALP production tensor can be calculated perturbatively using the effective interaction of Eq.~\eqref{eq:Leff}, which gives\footnote{This formalism is well known for the case of electron scattering, with the trivial replacements $c_\gamma/\Lambda\to e$, and
$L^{\mu\nu}=2 \left( k^\mu p^\nu + k^\nu p^\mu - g^{\mu\nu} (k\cdot p-m^2_e) \right)\,$, see Ref.~\cite{Donnelly:2017aaa} for further details.}
\begin{align}
	L^{\mu\nu}
	=	 -\frac{1}{8} \left( t - m^2_a \right)^2 g^{\mu\nu}
	-\left( \frac{t - m_a^2}{4}\right) ( k_\gamma^\mu p_a^\nu +  k_\gamma^\nu p_a^\mu)
	- \frac{m^2_a}{2} k_\gamma^\nu k_\gamma^\mu    \, .
\end{align}
Finally, the double differential cross section is given by
\begin{align}
	\frac{ d^2 \sigma }{d m^2_X \, dt}
	=	\frac{-t}{2(t+m_N^2-m_X^2)^2} \frac{ d^2 \sigma }{dx \, dt}
	=	\frac{1}{ (s-m_N^2)^2}\frac{\abs{\overline{\cM}}^2}{16\pi}\, .
\end{align}
The integration boundaries are
\begin{align}
	\label{eq:tlimits}
	m^2_X \in \left[ m^2_N , (\sqrt{s} -m_a)^2 \right] \, , \qquad
	t_0 \, (t_1)
	= 	\left[ \frac{ m^2_X -m^2_a - m^2_N  }{2\sqrt{s}} \right]^2 - \left( k_{\rm cm} \mp p_{\rm cm} \right)^2 \, ,
\end{align}
with
\begin{align}
	\label{eq:kpcm}
	k_{\rm cm}
	=	\frac{k m_N}{\sqrt{s}} \, , \qquad
	p_{\rm cm}
	=	\sqrt{ \left(  \frac{s+ m^2_a - m^2_X}{2\sqrt{s}} \right)^2- m^2_a} \, .
\end{align}

We estimate the nuclear response functions $\widetilde{W}_{1,2}$ in terms of the single nucleon response functions $W_{1,2}$ using the relativistic Fermi gas model as in Ref.~\cite{Alberico:1988bv} and the normalization of Ref.~\cite{Donnelly:2017aaa}:
\begin{align}
	W_T = m_N R_T  = -\left(g_{ij}+ \frac{q_i q_j}{|\vec{q}|^2}\right)W^{ij}\,,\quad\quad
	W_L = m_N R_L  = W^{00}\,.
\end{align}
The transverse and longitudinal response functions are related to those of Eq.~\eqref{eq:Hadronic tensor} via %
\begin{align}
	W_T  = 2\widetilde{W}_1\,,\qquad
	W_L  = -\widetilde{W}_1 \left(1 - \frac{t}{4 x^2 m_N^2}\right) + \widetilde{W}_2 \left(1 - \frac{t}{4 x^2 m_N^2}\right)^2 \,. 
\end{align}
By using the above definition of $\rho$ and inverting these relations we find
\begin{align}
	\widetilde{W}_1  =\frac{1}{2} W_T \,,\quad\quad
	\widetilde{W}_2  = \rho^2 W_L  +\frac{\rho}{2} W_T   \,.
\end{align}
The above equation should be understood to be the sum of the neutron and proton response functions,
\begin{align}
	W_{T,L} = W_{T,L}^{(n)} + W_{T,L}^{(p)} \,.
\end{align}
The explicit forms of the longitudinal and transverse response functions are
\begin{align}
	W^{(a)}_{L,T}
= 	&\frac{3m_N\mathcal{N}^{(a)}}{4 m_p \kappa \eta_F^3}(\epsilon_F - \Gamma)\theta(\epsilon_F - \Gamma)
	\times
	\begin{cases}
		\frac{\kappa^2}{\tau} \left\{ \left[(1+\tau)W^{(a)}_2(\tau)-W^{(a)}_1(\tau)\right] + W^{(a)}_2(\tau)\Delta  \right\}  & \text{for L}\\
		2W^{(a)}_1(\tau)+W^{(a)}_2(\tau)\Delta & \text{for T}
	\end{cases} \,.
\end{align}
Above, the superscript $a = p\,(n)$ stands for protons\,(neutrons) in the gas, $m_p\approx m_n$ is the proton\,(neutron) mass, $\mathcal{N}^{(p)} = Z$, and $\mathcal{N}^{(n)} = A-Z$.
We also define the dimensionless parameters
\begin{align}
	\kappa & = \frac{|\vec{q}|}{2 m_p }= \frac{1}{2m_p} \sqrt{-\frac{t}{\rho}} \,, \\
	\lambda  &= \frac{\omega}{2 m_p }= - \frac{t}{4 x m_p m_N}\,, \\
	\tau & = \kappa^2 - \lambda^2 = -\frac{t}{4m^2_p} \, , \\
	\eta_F & = \frac{p_F}{m_p} \,, \\
	\varepsilon_F &= \sqrt{1+\eta_F^2} \,,
\end{align}
where $p_F \approx m_p/4 \approx 235\,$MeV is the Fermi momentum~\cite{Alberico:1988bv} and
\begin{align}
	\Gamma & = \max \left[\varepsilon_F - 2\lambda ,\, \kappa\sqrt{1+\frac{1}{\tau}}-\lambda\right] \,,\\
	\Delta & = \frac{\tau}{\kappa^2}\left[\frac{1}{3}\left(\varepsilon_F^2 +\varepsilon_F \Gamma +\Gamma^2\right) + \lambda\left(\varepsilon_F + \Gamma\right) + \lambda^2 \right] - (1+\tau)~.
\end{align}
Finally, the single nucleon response functions are related to the Sachs form factors Eqs.~\eqref{eq:Proton form factors E}--\eqref{eq:Proton form factors M} by~\cite{Alberico:1988bv}
\begin{align}
	W^{(a)}_1(\tau) = \tau \left[G^{(a)}_{\rm M}(\tau)\right]^2  \,, \quad\quad
	W^{(a)}_2(\tau) = \frac{1}{1+\tau}\left\{ \left[G^{(a)}_{\rm E}(\tau) \right]^2+\tau \left[G^{(a)}_{\rm M}(\tau)\right]^2\right\}\,,
\end{align}
where the Sachs form factors are given by~\cite{RevModPhys.29.144,PhysRevLett.8.110}
\begin{align}
&	G^p_{\rm E}(t) =  \frac{1}{ \left( 1- t/q^2_0 \right)^2 } \, , \label{eq:Proton form factors E} \\
&	G^p_{\rm M}(t) = \frac{\mu_p}{ \left( 1- t/q^2_0 \right)^2 } \, , \label{eq:Proton form factors M} \\
&	G^n_{\rm E}(t) =  \frac{-t}{4m^2_p}\frac{\mu_n}{ \left( 1- t/q^2_0 \right)^2 } \frac{1}{1-5.6t/m^2_p} \approx 0 \, , \label{eq:Neutron form factors E}  \\
&	G^n_{\rm M}(t) =  \frac{\mu_n}{ \left( 1- t/q^2_0 \right)^2 } \, , \label{eq:Neutron form factors M}
\end{align}
with $\mu_p = 2.793\,(\mu_n =-1.913 )$~\cite{PDG} the magnetic moment of the proton\,(neutron) and $q^2_0 = 0.71\, {\rm GeV}^2$.
The resulting QE cross section is given by
\begin{align}
	\label{eq:XsecNQE}
	\frac{d^2\sigma^{\rm QE}_{\gamma N \to a X}}{dt \, d x}
	& =	\frac{ \alpha g^2_{a\gamma} }{32\Lambda^2}
	\frac{ w_1 \widetilde{W}_1 + w_2 \widetilde{W}_2}{m^2_N t x^2 (m_N^2 - s)^2} \, ,
\end{align}
where
\begin{align}
	w_1
	=&	2m^2_N (m_a^2 - t)^2 \, , \\
	w_2
	=&  	 m_N^2 m_a^2( 2 t - m_a^2) - t \left[(s-m_N^2)^2+t m_N^2\right] - \frac{t}{x}(s-m_N^2)(t-m_a^2) \,.
\end{align}

\subsection{Elastic scattering limit}

Elastic scattering corresponds to the following limit (see discussion in Ref.~\cite{Donnelly:2017aaa}):
\begin{align}
	m_X^2 = m_N^2
	\quad \Rightarrow \quad
	\omega = -\frac{t}{2m_N}
	\quad\Rightarrow\quad
	x = 1 \, .
\end{align}
The nuclear response function is obtained by integrating Eq.~\eqref{eq:Hadronic tensor}, which gives
\begin{align}
	F^{\rm had}_{\mu\nu}
	=	16\pi m^2_N \left( -F^2_1\left(g^{\mu\nu} - \frac{q^\mu q^\nu}{t} \right) + F^2_2 V_i^\mu V_i^\nu \right) \, ,
\end{align}
and
\begin{align}
	F^2_1 = \frac{F^2_T}{2} \, , \quad\quad
	F^2_2 = \frac{\rho}{2}\left( F^2_T + 2\rho F^2_L \right) \, .
\end{align}
The transverse and longitudinal response functions are approximately
\begin{align}
	F^2_T (t) \simeq 0 \, , \quad\quad
	F^2_L (t) \simeq \frac{Z^2}{4\pi} F^2_0 (t) \, ,
\end{align}
where
\begin{align}
	F_0(t)
	= \frac{1}{Z} \int d^3x \frac{\sin(\abs{\vec{q}}x) }{\abs{\vec{q}}x} \rho(x)_{00} \, , \quad\quad
	\rho(x)_{00}
	=	\frac{\rho_0}{ 1+ \exp\left( \frac{x-R}{a}\right)} \, .
\end{align}
Above, $\abs{\vec{q}} = \sqrt{t(t/4m^2_N-1)}$, $a$ and $R$ are determined from data using the two parameter Fermi model (see Ref.~\cite{DeJager:1974liz}).
Finally, $\rho_0$ is defined by the normalization condition $\int d^3 x \rho(x)_{00} = Z\,$.
Therefore, the amplitude squared is
\begin{align}
	\abs{\overline{\cM}}^2
	=&	2\pi\frac{8e^2 g_{\rm lep}^2}{ t^2}\rho^2 L^{\mu\nu}   \left( k_N^\mu + \frac{q^\mu}{2} \right)\left( k_N^\nu + \frac{q^\nu}{2} \right)
	\frac{Z^2}{4\pi} F^2_0 (q) \, ,
\end{align}
and the elastic cross section is
\begin{align}
	\frac{d\sigma}{dt}
	=& 	\frac{\abs{\overline\cM}^2}{64\pi s k^2_{{\rm cm}}}  \, ,
\end{align}
where the integration boundaries are given in Eq.~\eqref{eq:tlimits} for $m_X= m_N$ and $k_{\rm cm}$ is defined in Eq.~\eqref{eq:kpcm}.
The resulting elastic cross section is
\begin{align}
	\label{eq:XsecNEl}
	\frac{d\sigma^{\rm El}_{\gamma N \to a N}}{dt}
	& =	2\alpha Z^2  \frac{g_{a\gamma}^2}{\Lambda^2}
	\frac{ m_N^4  \left(m_a^2 t \left(m_N^2+s\right) - m_a^4 m_N^2 - t \left(\left(m_N^2-s\right)^2+s t\right)\right)}{ t^2 \left(m_N^2-s\right)^2 \left(t-4 m_N^2\right)^2}
	F^2_0(q) \, .
\end{align}

\subsection{Proton target}

A simple case is a proton target, where we consider elastic scattering.
The proton effective vertex is
\begin{align}
	\label{eq:LpFF}
	\cL_{{\rm eff}-p}
	=	e\bar{\psi}_p\left( \gamma_\mu F^p_1(t) + \frac{i\sigma_{\mu\nu}}{2m_p} F^p_2(t) q^\nu \right) \psi_p \, ,
\end{align}
where
\begin{align}
	&	F^p_1(t) = \frac{4 m^2_p G^p_{\rm E}(t) - t G^p_{\rm M}(t) }{4m^2_p - t} \, , \quad
	F^p_2(t) = \frac{4 m^2_p \left( G^p_{\rm M}(t) - G^p_{\rm E}(t) \right) }{4m^2_p - t} \, .
\end{align}
and $G^p_{\rm E,M}$ are given in Eqs.~\eqref{eq:Proton form factors E} and~\eqref{eq:Proton form factors M}.
The cross section is
\begin{align}
	\label{eq:dsigaa}
	\frac{d\sigma_{\gamma p \to a p}}{dt}
	& =	\frac{\alpha g_{a\gamma}^2}{32 \Lambda^2}\frac{f_1 F^{p,2}_1(t)+f_{12} F^p_1(t) F^p_2(t) + f_2 F^{p,2}_2(t)}{m_p^2 \, t^2 \, (m_p^2-s)^2} \, ,
\end{align}
with
\begin{align}
	\label{eq:f1}
	f_1
	& =  	2 m_p^2 \left\{ 2 m_a^2 t (m_p^2 + s + t) - m_a^4 (2 m_p^2 + t)  - t \left[2 (m_p^2 - s)^2 + 2 s t + t^2\right] \right\} \, , \\
	\label{eq:f2}
	f_2
	& = 	t \left\{ t \right[ m_p^4 - 2 m_p^2 (s + t) + s (s + t) \left]  + t \, m_a^2 (3 m_p^2 - s) - m_a^4 m_p^2  \right\} \, , \\
	\label{eq:f12}
	f_{12}
	& = 	4 m_p^2 \, t \, (m_a^2 - t)^2 \, .
\end{align}

\subsection{Quasi-Elastic data-driven signal estimation}
The ALP Primakoff signal is proportional to the Primakoff production yield of $\pi^0$, $\eta$ or $\eta'$.
Therefore, by measuring these yields for different $s$, $t$, and $x$ one can preform data-driven signal estimation and normalize the ALP search to the data without relying on theory inputs for the form factors.
In addition, the experimental photon flux does not need to be know, and only the response of the detector at $m_a$ relative to $m_P$ is required.
Here, we discuss the general case of quasi-elastic scattering on a heavy nucleus.
The elastic case is discussed in the Letter.
In principle, this method can also be used to preform a measurement of the $\Gamma_{\eta^{(\prime)}\to\gamma\gamma}/\Gamma_{\pi^0\to\gamma\gamma}$ ratio with reduced hadronic uncertainties.

For a given target, $N$, the ratio between ALP and $P=\pi^0, \eta, \eta'$ Primakoff production is
\begin{align}
	\label{eq:Rap}
	\cR_{aP}(s,t, x, m_a,m_P)
	=&	\left( \frac{d\sigma_{\gamma N \to a X}}{dt d x}\right) \Big/ \left( \frac{d\sigma_{\gamma N \to P X}}{dt d x} \right) \nonumber\\
	=&	\frac{  g^2_{a\gamma}/\Lambda^2 }{64\pi \Gamma_{P\to\gamma\gamma}/m^3_P} \frac{(m^2_a - t)^2}{(m^2_P - t)^2}
	\frac{  1 + w_{21}(s,t, x, m_a) W_{21}(t,x)}{ 1 + w_{21}(s,t, x, m_P) W_{21}(t,x)} \, ,
\end{align}
with
\begin{align}
	W_{21}(t,x)
	= 	\widetilde{W}_2(t,x)/\widetilde{W}_1(t,x) \, , \quad\quad
	w_{21}(s,t, x, m_a)
	=	w_2 / w_1  \, .
\end{align}
From Eq.~\eqref{eq:Rap} we see that in order to normalize ALP production to pseudoscalar production one needs to know $W_{21}\,$.
This can be achieved by measuring the differential cross section at two different $s$ values:
\begin{align}
	\cR_{PP}(s_1,s_2,t,x,m_P)
	=&	\left( \frac{d\sigma_{\gamma N \to P X}}{dt d x} \right)_{s=s_1} \Big/ \left( \frac{d\sigma_{\gamma N \to P X}}{dt d x} \right)_{s=s_2} \nonumber \\
	=&	\frac{ 1 + w_{21}(s_1,t, x, m_P) W_{21}(t,x)}{ 1 + w_{21}(s_2,t, x, m_P) W_{21}(t,x)}
	\frac{ (m_N^2 - s_2)^2}{ (m_N^2 - s_1)^2} \, ,
\end{align}
which allows us to determine $W_{21}(t,x)\,$ and estimate the ALP Primakoff production cross section.
We note that this should be done on the data around the Primakoff peak where strong pseudoscalar-meson production can be neglected.

\section{ALP strong photoproduction}

In the Letter, we used an approximation to obtain the data-driven normalization for strong ALP photoproduction.
This was because, as discussed in Ref.~\cite{AlGhoul:2017nbp}, $\eta$ production at \gluex energies, while clearly dominantly $t$-channel, is not yet fully understood.
Here, we show that---once both $\pi^0$ and $\eta$ photoproduction are well understood---it is possible to derive a fully data-driven normalization strategy similar to the one we proposed above for Primakoff production.

We estimate this contribution using the vector-meson-dominance (VMD) model.
The process begins with an insertion of $\gamma$--$\rho$\,($\gamma$--$\omega)$ mixing, an interaction of the form $PVV^{(\prime)}$ where $P=\pi,\,\eta,\,\eta^\prime$ or $P=a$ for the ALP and $V=\rho,\,\omega$ (see Ref.~\cite{Fujiwara:1984mp} and the Supplemental Material of Ref.~\cite{Aloni:2018vki} for further details).
The ALP-gluon coupling can be replaced by an ALP $U(3)$ representation by performing a chiral transformation of the light-quark fields\,\cite{Georgi:1986df,Bardeen:1986yb,Krauss:1986bq} .
Following Ref.~\cite{Aloni:2018vki}, we denote this as
$\boldsymbol{a}$ and take this $m_a$-dependent representation directly from Ref.~\cite{Aloni:2018vki}.

For the photon-meson mixing we define
\begin{align}
	\gamma_\rho = 3\gamma_\omega = e/g \approx 4.9\times 10^{-2} \, ,
\end{align}
where $g\approx \sqrt{12\pi}$ is the VMD coupling~\cite{Fujiwara:1984mp}.
As this factor is already at the percent level, we do not consider more than a single insertion.
For the $\rho$ and $\omega$ couplings to the proton, we follow Ref.~\cite{Machleidt:1987hj}.
Comparing to our notations of the $F_{1,2}$ form-factors in Eq.~\eqref{eq:LpFF} we have
\begin{align}
	&e F_1^\rho \to g_{\rho NN} \approx 3.25  \, ,\\
	&e F_2^\rho \to g_{\rho NN}  \kappa_{\rho NN} \approx 20.7 \,,\\
	&e F_1^\omega \to g_{\omega NN} \approx 15.9 \, ,\\
	&e F_2^\omega\to g_{\omega NN}  \kappa_{\omega NN}\approx 0 \, .
\end{align}

The $\gamma p \to a p$ cross section from $\gamma$--$V_{1}$ mixing and $t$-channel exchange of $V_{2}$, including the interference with $\gamma$--$V_{1^\prime}$ mixing and $t$-channel exchange of $V_{2^\prime}$, is given by
	\begin{align}
		\label{eq:Full_strong_cross_section}
	\frac{d\sigma}{dt}
	=& 	\sum_{V_{i}=\rho,\,\omega} \frac{\gamma_{V_1}\gamma_{V_{1'}} g_{a V_1 V_2}g_{a V_{1'} V_{2'}}g_{V_2 NN} g_{V_{2'} NN} }{128\pi}
	\cP_{V_2} \cP^*_{V_{2'}}
	\frac{f_1  + f_{12} \left(\kappa_{V_{2'} NN} + \kappa_{V_2 NN}\right)/2 + f_2 \kappa_{V_2 NN}\kappa_{V_{2'} NN}}
	{m_p^2 (m_p^2-s)^2} \, ,
	\end{align}
where $f_{1,2,12}$ are given in Eqs.~\eqref{eq:f1}--\eqref{eq:f12}.
The Regge propagator, $\cP_{V}(t)$, replacing the vector meson Breit-Wigner propagator is
\begin{align}
	\cP_{V}(t)
= 	\frac{\pi}{ \Gamma(\alpha(t))} \frac{ 1 - e^{-i\pi \alpha(t)} }{2\sin(\pi\alpha(t))} \left( \frac{s}{\rm GeV^2}\right)^{\alpha(t)-1} \, ,
\end{align}
with $\alpha_V(t) = \alpha_V^0 + \alpha^\prime_V t$, with $\alpha^0_V = 0.5\,$,  $\alpha^\prime_V =0.9\, {\rm GeV^{-2}}$~\cite{Mathieu:2015eia}.
Finally, the ALP--vector-meson coupling in Eq.~\eqref{eq:Full_strong_cross_section} is given by
\begin{align}
	g_{a V_1 V_2} =  -\frac{3 \, g^2 }{8\pi^2 f_a}  {\ensuremath{\langle \boldsymbol{a V_1 V_2} \rangle}} \,,
\end{align}
where $ {\ensuremath{\langle \boldsymbol{a V_1 V_2} \rangle}\xspace}$ is the ALP-vector-vector effective interaction,  see Ref.~\cite{Aloni:2018vki}
(for a pseudoscalar meson, $\boldsymbol{a}$ is replaced by the appropriate $U(3)$ representation, and $f_a \to f_{\pi}$).
Under the assumption that this model adequately describes both $\pi^0$ and $\eta$ production, it is straightforward to obtain the ratio of the ALP and $P=\pi^0,\eta$ differential cross sections from the equations above and perform a fully data-driven normalization similar to what we proposed for Primakoff production.
Similarly, this same procedure can be followed for any other pseudoscalar-meson photoproduction model.

To estimate the accuracy of the approximation made in the Letter, we assume that the model described above is valid for both $\pi^0$ and $\eta$ photoproduction.
To be conservative, we allow for an arbitrary phase between the $\rho$ and $\omega$ exchange amplitudes, and choose its value to maximize the discrepancy induced by the approximation used in the Letter.
We find that the largest error is less than a factor of 2, and as expected, occurs near 0.4\,GeV where $\langle \boldsymbol{a \pi^0} \rangle$ and $\langle \boldsymbol{a \eta} \rangle$ are roughly equal in size.


\section{PrimEx details}

This section describes the \primex experiment and data-taking conditions, along with the details of our toy \primex Monte Carlo simulation and our bump hunt of the published $m_{\gamma\gamma}$ spectrum from one angular bin of carbon-target data in Ref.~\cite{Larin:2010kq}.

\subsection{\primex Description}

The first run of the \primex experiment was in Hall~B at Jefferson Lab in 2004~\cite{Larin:2010kq}.
Data were collected on both C and Pb targets using a 4.9--5.5\,GeV photon beam and a high-resolution multichannel calorimeter.
The integrated luminosities were 0.3/pb for C and 0.002/pb for Pb.
A follow-up run of \primex was performed in 2010, which collected 0.4/pb on C and 0.3/pb on Si.
The target and beam properties for each run are given in Table~\ref{tab:primex}.
The \primex calorimeter was located 7.5\,m downstream of the targets with an acceptance covering $\approx 1$\,m$^2$.
A 4.1\,cm$\times$4.1\,cm square hole was left at the center to allow the beam to pass through.
This calorimeter geometry provides good acceptance for $0.03 \lesssim m_a \lesssim 0.3$\,GeV.
We developed a simple toy Monte Carlo of this set up, including the acceptance, efficiency, and resolution of the calorimeter taken from Refs.~\cite{kubantsev2006performance,larin2011new}.

\begin{figure}[t]
\includegraphics[width=0.7\textwidth]{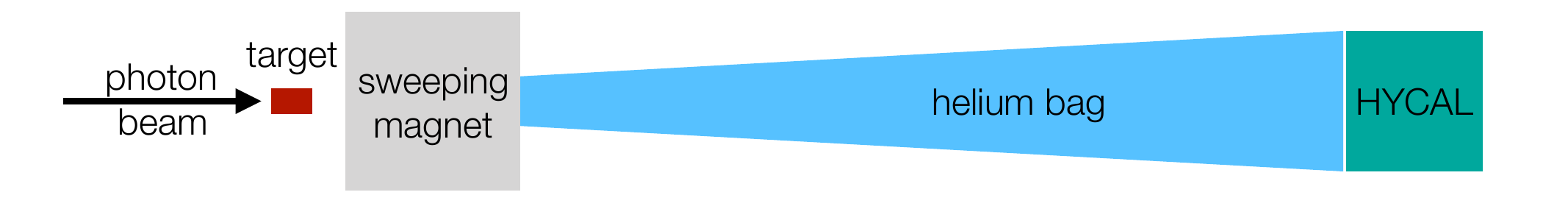}
\caption{
Schematic of the \primex experiment (not to scale) including the high-resolution multichannel calorimeter (HYCAL) which is vital to the search proposed here.
 }
\label{fig:primex_diagram}
\end{figure}

\begin{table}
\begin{tabular}{ccccc}
  \qquad \primex run \qquad & \qquad target \qquad & \qquad thickness [$X_0$] \qquad & \qquad  integrated luminosity [1/pb] \qquad  & \qquad  beam energy [GeV] \qquad \\
\hline
I & $^{12}$C & 5\% & 0.3 & 4.9--5.5 \\
I & $^{208}$Pb & 5\% & 0.002 & 4.9--5.5 \\
II & $^{12}$C & 8\% & 0.4 & 4.4--5.3 \\
II & $^{28}$Si & 10\% & 0.3 & 4.4--5.3 \\
\hline
\end{tabular}
\caption{\primex data-taking conditions in both experimental runs.}
\label{tab:primex}
\end{table}

Ref.~\cite{Larin:2010kq} used the following selection to measure $\Gamma_{\pi^0\to\gamma\gamma}$ (we adopt this selection for the ALP search):
the incident photon energy had to be in the range 4.9--5.5\,GeV;
the sum of the two decay photon energies was required to be larger than 2.5\,GeV;
and the {\em elasticity}, defined as the sum of the two decay photon energies divided by the photon beam energy, had to be within 3 times the resolution of unity.
The product of the acceptance and efficiency is $\gtrsim 30\%$ in the region $0.03 < m_a < 0.3$\,GeV, but drops quickly outside of this region due to the acceptance.
We find that the resolution should be $\approx 0.02 \times m_a$ in this region, which agrees with the value of $0.017 \times m_{\pi^0}$ found in Ref.~\cite{Larin:2010kq}.

To estimate the sensitivity of each \primex data sample, we need to determine the mass dependence of the efficiency and to estimate the background versus $m_{\gamma\gamma}$ in each sample.
Using the known nuclear form factors and Primakoff differential cross section~\cite{Donnelly:2017aaa,Alberico:1988bv,DeJager:1974liz}, we generate Primakoff $\pi^0$ Monte Carlo events for the \primex photon-beam energy.
We require that both photons from the $\pi^0\to\gamma\gamma$ decay are in the \primex calorimeter fiducial acceptance region and apply the required smearing to account for resolution.
We then apply the full selection of Ref.~\cite{Larin:2010kq} and find that our predicted $\pi^0$ Primakoff yields are consistent with those observed in Refs.~\cite{PrimExPAC33,YangThesis}.
We assume that the reconstruction efficiency is independent of the ALP mass in the search region (excluding geometrical acceptance effects).
In addition, we assume that the ALP bump hunt will only use candidates with $\theta_{\gamma\gamma} < 0.5^{\circ}$, where $\pi^0$ production is dominated by the Primakoff process for all targets.
We determine the efficiency at each $m_a$ using our toy Monte Carlo and these selection criteria.

Obtaining a data-driven background estimate will be straightforward for the \primex collaboration using the $m_{\gamma\gamma}$ sidebands at each $m_a$ (see, {\em e.g.}, Refs.~\cite{Williams:2015xfa,Williams:2017gwf}).
However, estimating the background for this study---without access to the data---is considerably more difficult.
We considered many possible backgrounds, {\em e.g.}\ $\gamma N \to N \omega(\pi^0[\gamma\gamma]\gamma)$ where one photon is not reconstructed or the $\pi^0$ photons are merged into a single cluster, though we found that no hadronic reactions are capable of contributing background at a rate comparable to that observed in Fig.~2 of Ref.~\cite{Larin:2010kq}.
Therefore, we conclude that the \primex background is dominantly due to electromagnetic interactions of the photon beam with the target that produce either additional photons or $e^+e^-$ pairs.
Figure~2 of Ref.~\cite{Larin:2010kq} shows the forward-most angular region.
Given that the beam backgrounds should decrease moving away from the beam line, using this angular bin---and assuming a uniform $\theta_{\gamma\gamma}$ distribution---provides a conservative background estimate.
We also conservatively assume that the backgrounds above (below) the $m_{\gamma\gamma}$ region shown in Fig.~2 take on the values at the upper (lower) edge of the plot, even though they appear to be decreasing in both directions (which is expected).
The resulting predicted background---excluding the $\pi^0$ peak region---is between 40 and 300 candidates per $\pm 2\sigma$ window for the first \primex carbon-target run.
We scale the beam-induced background, which is shown for the first C run, by the product of the target radiation length and the number of photons on target for each \primex run (see Table~\ref{tab:primex}).

\subsection{Bump Hunt of Fig.~2 of Ref.~\cite{Larin:2010kq}}

\begin{figure}[!t]
\includegraphics[width=0.57\textwidth]{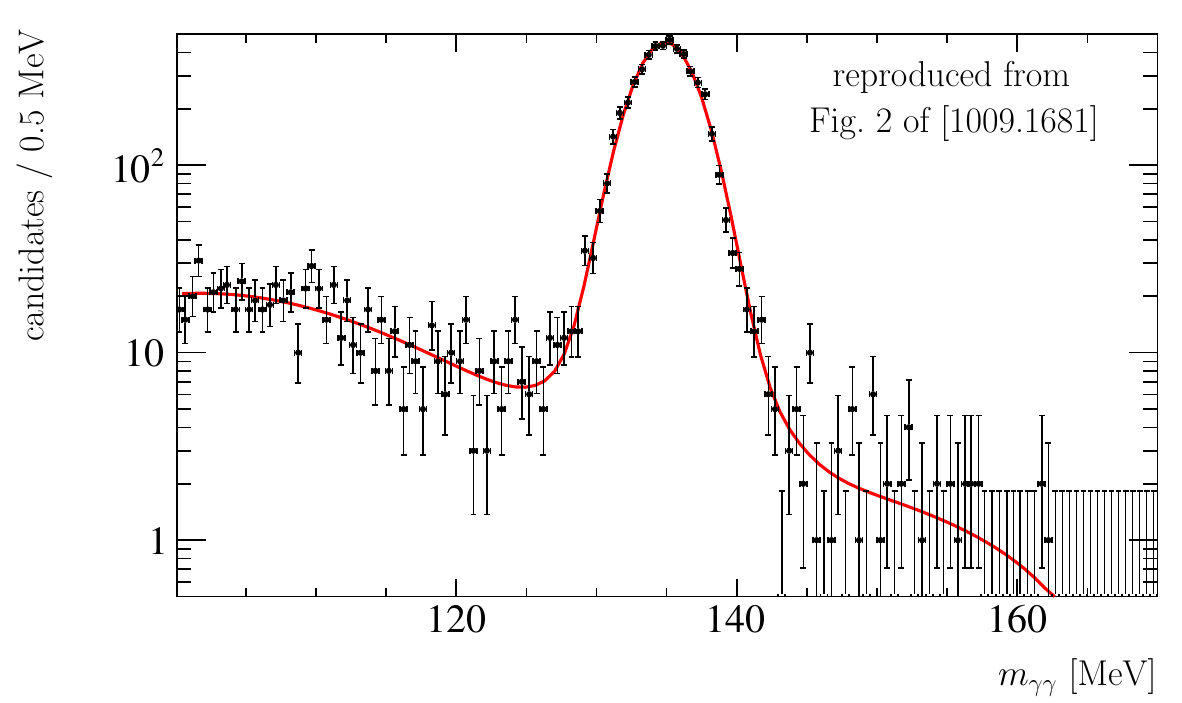}
\includegraphics[width=0.35\textwidth]{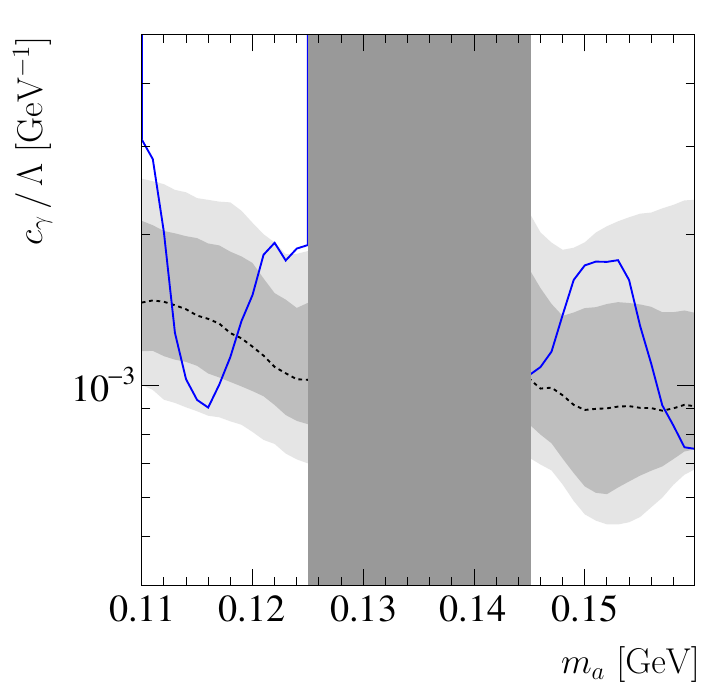}
\caption{
(left) Reproduction of Fig.~2 of Ref.~\cite{Larin:2010kq} fit to the background-only model described in the text.
(right) Limits obtained from (solid blue) our bump hunt of this $m_{\gamma\gamma}$ distribution compared with the (dashed) expected, (dark shaded) $\pm 1\sigma$, and (light shaded) $\pm 2\sigma$ regions.
 }
\label{fig:primex}
\end{figure}

Ref.~\cite{Larin:2010kq} published the diphoton mass spectrum near the $\pi^0$ peak for one forward angular bin from the C data obtained in the first \primex run (see Fig.~2 of Ref.~\cite{Larin:2010kq}).
We have digitized this plot (see Fig.~\ref{fig:primex}).
We scan the $m_{\gamma\gamma}$ spectrum in steps of roughly half the resolution (1~MeV steps).
At each mass, a binned maximum likelihood fit is performed to the data in Fig.~\ref{fig:primex}.
The profile likelihood is used to determine the confidence interval of the number of $a \to \gamma\gamma$ decays observed, from which an upper limit at 95\% confidence level is obtained.
The confidence intervals are defined using the {\em bounded likelihood} approach, which involves taking the change in the likelihood relative to zero signal, rather than the best-fit value, if the best-fit signal value is negative.
This enforces that only physical (nonnegative) upper limits are placed on the ALP yield, and prevents defining exclusion regions that are much better than the experimental sensitivity in cases where a large deficit in the background yield is observed.

The fit model contains contributions from $\pi^0 \to \gamma\gamma$, combinatorial diphoton combinations, a broad peak-like structure below the $\pi^0$ mass, and an $a \to \gamma\gamma$ signal component.
The $\pi^0$ contribution is described by a double Gaussian function, where both Gaussians share the same mean.
The combinatorial background is modeled by the {\em ad hoc} function $(c_0 + c_1 m_{\gamma\gamma})\times (1-{\rm exp}\{c_2 (m_{\gamma\gamma}-c_3)\})$ (the $c_i$ are free parameters in the fit), which is observed to describe the data well.
There is a broad peak-like structure in the background below the $\pi^0$ mass.
We model this contribution using a single wide Gaussian.
Finally, the signal is modeled using the same PDF as that of the $\pi^0$ except, of course, for the value of the mean mass which is fixed for each test mass value.
Figure~\ref{fig:primex} shows the $m_{\gamma\gamma}$ spectrum fit to the background-only model, which describes the data well.
The observed $\pi^0$ yield is $\approx 5100$.
The constraints on $c_{\gamma}/\Lambda$ are obtained from the upper limit on the number of $a \to \gamma\gamma$ decays observed at each $m_a$ as described above.
These are shown in Fig.~\ref{fig:primex} compared to their expected values.
Near 115 and 160\,MeV, these are the best limits set to date on the ALP-photon coupling.
The sensitivity using the full \primex data set---rather than just one angular region of the C data---is expected to be substantially better, and will cover a much larger mass region.

\section{GlueX details}

This section describes the \gluex experiment and data-taking conditions, along with the details of our toy \gluex Monte Carlo simulation and our bump hunt of the published $m_{\gamma\gamma}$ spectrum from $\approx 1$/pb of proton-target data in Ref.~\cite{AlGhoul:2017nbp}.

\subsection{\gluex Description}

The \gluex experiment began taking data in Hall~D at Jefferson Lab in 2016. In the spring 2016, $\approx 1$/pb of data on a liquid-hydrogen target were collected with a linearly polarized 8.2--9.2\,GeV photon beam.
These data were used to measure the beam asymmetry $\Sigma$ for both $\pi^0$ and $\eta$ photoproduction~\cite{AlGhoul:2017nbp}.
The experiment has collected $\approx 50$/pb thus far, and plans to collect $\mathcal{O}(1/{\rm fb})$ of data using its nominal liquid-hydrogen target~\cite{Dugger:2012qra}.
\gluex has both forward and central calorimeters.
We take the \gluex acceptance, efficiency, and resolution from Refs.~\cite{beattie2018construction,hardin2018upgrading}.
The most important component of the \gluex detector to our studies is the forward calorimeter, which is located about 5.6\,m downstream of the target and
covers roughly from 2 to 11$^{\circ}$ in the lab frame.
Its resolution is roughly $3.5\% + 5.7\% / \sqrt{E_{\gamma} / {\rm GeV}}$.

An updated version of the \primex experiment recently ran using the \gluex detector with an additional small-angle calorimeter~\cite{Primakoff2010}.
This new experiment used a helium target, which makes it less sensitive than \primex for ALPs; however,
several proposals have been made for future \gluex running with heavy nuclear targets~\cite{PAC}.
Specifically, we consider a Pb target here, though other targets are possible and it is simple to rescale our results for other nuclei.
We take the acceptance, efficiency, and resolution for the small-angle calorimeter from Ref.~\cite{Primakoff2010}.
The small-angle calorimeter is located about 4\,m downstream of the nominal \gluex forward calorimeter,
and fully covers the acceptance hole in the nominal \gluex forward calorimeter.
As was done for \primex, a 4.1\,cm$\times$4.1\,cm square hole was left at the center to allow the beam to pass through.

\begin{figure}[b]
\includegraphics[width=0.7\textwidth]{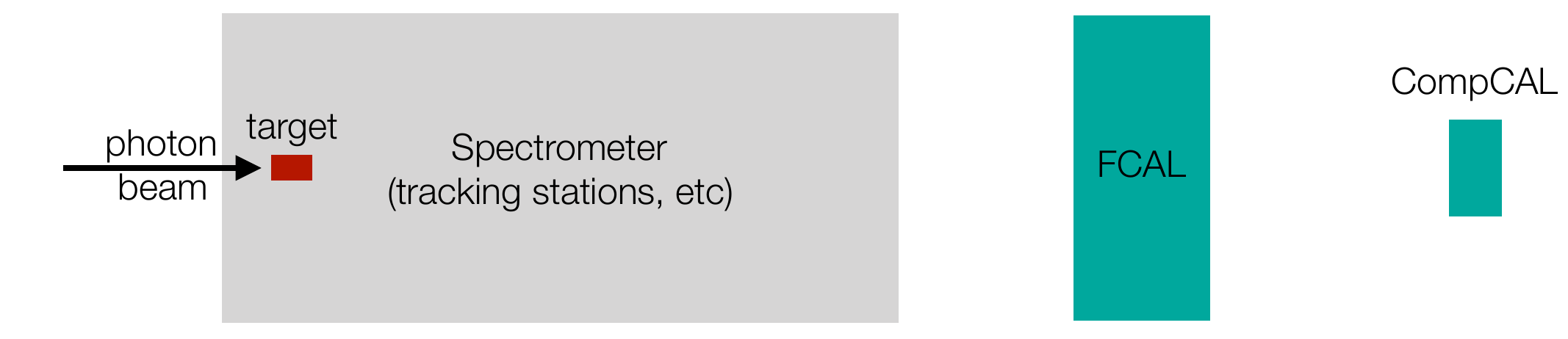}
\caption{
Schematic of the \gluex experiment (not to scale) including the forward (FCAL) and an additional small-angle (CompCAL) calorimeters,
which are vital to the search proposed here.
 }
\label{fig:gluex_diagram}
\end{figure}

For $m_a < m_{\eta}$, we rescale the expected beam background from the \primex Pb run.
There are three additional backgrounds that contribute to the \gluex run at higher masses:
Primakoff production of $\eta$ and $\eta'$ mesons, and
coherent nuclear production of $\gamma N \to N \omega(\pi^0[\gamma\gamma]\gamma)$.
The cross sections for these processes are well known, making it straightforward to estimate their yields using Monte Carlo.
That said, these backgrounds are peaking, so we exclude the $\eta$ and $\eta'$ regions, and apply a relative 1\% systematic uncertainty on the background yield between the $\eta$ and $\eta'$ masses.
This systematic is the dominant effect in our prediction of the sensitivity in this mass region (this can likely be reduced in an actual search, improving the sensitivity greatly).
The \gluex experiment could explore regions of ALP parameter space where the ALP flight distance becomes nonnegligible.
Using Monte Carlo, we estimate that the impact on the ALP mass resolution and acceptance is small provided that its lab-frame flight distance is $\lesssim 30$\,cm (the length of the nominal liquid hydrogen target cell).
For simplicity, we apply a fiducial cut on the flight distance at 30\,cm, which is conservative since ALPs that decay after this distance could still be detected and a detailed study could determine the appropriate signal shape for each value of $\Lambda$.
We show projections for 2 luminosity values, though again it is easy to rescale these projections for other values.
The smaller data set corresponds to collecting 1/pb of Pb-target data.
The larger data set assumes that as much data is collected on a Pb target as is expected in the full \gluex proton-target run.
A more likely scenario would involve collecting a total nuclear-target data sample of roughly this size that consists of several smaller samples using different nuclei.
If this is the case, combining these samples using our data-driven approach is simple and the total sensitivity would be an $\mathcal{O}(1)$ factor worse than collecting only Pb data.

\subsection{Bump Hunt of Fig.~3 of Ref.~\cite{AlGhoul:2017nbp}}

\begin{figure}[!t]
\includegraphics[width=0.57\textwidth]{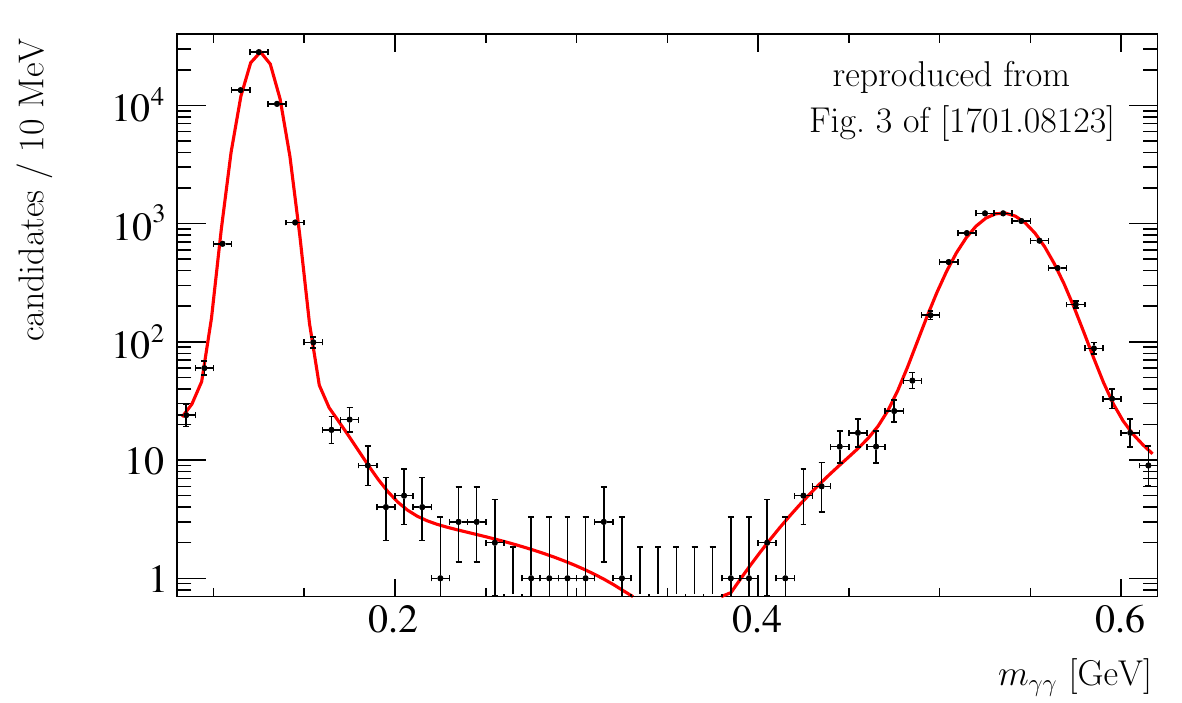}
\includegraphics[width=0.35\textwidth]{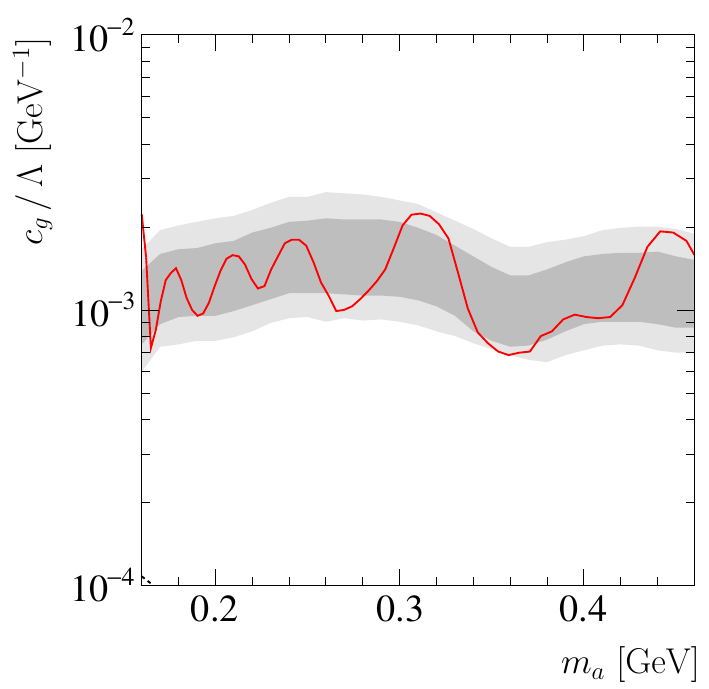}
\caption{
(left) Reproduction of Fig.~3 of Ref.~\cite{AlGhoul:2017nbp} fit to the background-only model described in the text.
(right) Limits obtained from (solid red) our bump hunt of this $m_{\gamma\gamma}$ distribution compared with the (dashed) expected, (dark shaded) $\pm 1\sigma$, and (light shaded) $\pm 2\sigma$ regions.
 }
\label{fig:gluex}
\end{figure}

Ref.~\cite{AlGhoul:2017nbp} published the $m_{\gamma\gamma}$ spectrum (see Fig.~3 of Ref.~\cite{AlGhoul:2017nbp}), along with the yields and efficiencies versus $t$ of both the $\pi^0$ and $\eta$ mesons (see Fig.~4 of Ref.~\cite{AlGhoul:2017nbp}).
We have digitized Fig.~3 (see Fig.~\ref{fig:gluex}).
We use this data to place constraints on $c_g / \Lambda$ using Eq.~\eqref{eq:strongyield}.
The ALP decay branching fraction is taken from Ref.~\cite{Aloni:2018vki}, though it is close to unity throughout this mass range.
Ref.~\cite{AlGhoul:2017nbp} provides the efficiencies versus $t$ at $m_{\pi^0}$ and $m_{\eta}$.
We discard the region $|t| < 0.1\,{\rm GeV}^2$ because the efficiency is small and sharply varying with $t$.
We discard the region $|t| > 1\,{\rm GeV}^2$ because our approximation in Eq.~\eqref{eq:strongyield} begins to break down here.
Note that, by necessity, we keep the background from these regions, since we have no way of removing it from the $m_{\gamma\gamma}$ spectrum of Fig.~3 in Ref.~\cite{AlGhoul:2017nbp}.
We linearly interpolate the efficiencies given at each $t$ for $m_{\pi^0}$ and $m_{\eta}$ to each $m_a$, and confirm this approach is valid to $\mathcal{O}(10\%)$ using toy Monte Carlo.
In this toy Monte Carlo, we generate the ALPs using the Regge model discussed above and the \gluex fiducial region described in the previous subsection; however, since the $[s,t]$ bins are small, the production model has negligible impact on obtaining the efficiencies.
Additionally, the same ALP lifetime correction is applied here as is applied for the Primakoff scenario, though this is a small correction in this case.

The approach used here is the same as for our \primex bump hunt.
We scan the $m_{\gamma\gamma}$ spectrum in steps of 10~MeV, which is the bin width of Fig.~3 of Ref.~\cite{AlGhoul:2017nbp}.
At each mass, a binned maximum likelihood fit is performed to the data in Fig.~\ref{fig:gluex}.
The profile likelihood is used to determine the upper limit on the number of $a \to \gamma\gamma$ decays observed using the bounded likelihood approach.
The fit model contains contributions from $\pi^0 \to \gamma\gamma$, $\eta\to\gamma\gamma$, combinatorial diphoton combinations, and an $a \to \gamma\gamma$ signal component.
The $\pi^0$ and $\eta$ contributions are described by double Gaussian functions.
The combinatorial background is modeled by a linear function.
The signal is modeled using a Gaussian PDF with a relative resolution fixed to the value observed for the $\eta$ peak of $\approx 3.5\% \times m_a$ (this constant relative resolution is confirmed by our toy Monte Carlo).

Figure~\ref{fig:gluex} shows the $m_{\gamma\gamma}$ spectrum fit to the background-only model, which describes the data well.
The constraints on $c_{g}/\Lambda$ are obtained from the upper limit on the number of $a \to \gamma\gamma$ decays observed at each $m_a$ as described above.
These are shown in Fig.~\ref{fig:gluex} compared to their expected values.
Over some of this mass range, these are the best limits set to date on the ALP-gluon coupling.
The sensitivity using the full \gluex data set---roughly 1000 times more luminosity---is expected to be about $1000^{\frac{1}{4}} \approx 5$ times better.

\end{document}